\documentstyle[aps,graphicx]{revtex}
\begin{document}
\begin{center}
{\bf Proton Drip-Line Calculations and the Rp-process}

\vspace{ 24pt}
B. A. Brown, R.R.C. Clement, H. Schatz and A. Volya\\
{\sl Department of Physics and Astronomy\\
and National Superconducting Cyclotron Laboratory,
Michigan State University, \\
East Lansing, Michigan 48824-1321, USA\\}

\vspace{ 24pt}
W. A. Richter\\
{\sl Department of Physics, University of Stellenbosch,\\
Stellenbosch 7600, South Africa}

\end{center}

\begin{abstract}

One-proton and two-proton separation energies are calculated for
proton-rich nuclei in the region $  A=41-75  $.
The method is based on Skyrme Hartree-Fock calculations of Coulomb
displacement energies of mirror nuclei in combination with the
experimental masses of the neutron-rich nuclei. The implications
for the proton drip line and the astrophysical rp-process are
discussed. This is done within the framework of a detailed analysis
of the sensitivity of rp 
process calculations in type I X-ray burst models on nuclear masses. 
We find that the remaining mass uncertainties, in particular for
some nuclei with $N=Z$, still lead to large
uncertainties in calculations of X-ray burst light curves. Further
experimental or theoretical improvements of nuclear mass data
are necessary before observed X-ray burst light curves can be used
to obtain quantitative constraints on ignition conditions and 
neutron star properties. We identify a list of nuclei for which 
improved mass data would be most important. 

\end{abstract}

\newpage

\section{Introduction}

The masses for the proton-rich nuclei above $  A=60  $ have not yet been
measured. However, they are important for the astrophysical
rapid-proton capture (rp) process \cite{WaW81}
which follows a path in nuclei near $  N=Z  $ for $  A=60-100  $. The rp
process is the dominant source of energy in type I X-ray bursts, and it
determines the crust composition of accreting neutron stars
\cite{WGI94,SAG97,SBC98,KHA99,SAB01}. It may also
be responsible for the p process nucleosynthesis of a few proton-rich 
stable nuclei in the $A=74$--$98$ mass range.
In
the absence
of experimental masses for the proton-rich nuclei, one often uses the
masses based upon the Audi-Wapstra extrapolation (AWE) method
\cite{b1}.
In this
paper we use the displacement-energy method
\cite{b2,b3,b4,b5}
to obtain the proton-rich masses with the Skyrme Hartree-Fock model
for the displacement energies.
The displacement
energy is the difference in the binding energies of
mirror nuclei for a given mass $  A  $ and isospin $  T  $:

\begin{equation}
D(A,T) = BE(A,T_{z}^{<}) - BE(A,T_{z}^{>}),
\end{equation}

\noindent
  where $  T=\,\mid T_{z}^{<}\mid\, =\,\mid T_{z}^{>}\mid   $, $
BE(A,T_{z}^{<})  $ is the binding energy of
the proton-rich nucleus and $  BE(A,T_{z}^{>})  $ is the binding energy
of the
neutron-rich nucleus. The displacement energy can be much more
accurately calculated than the individual $  BE  $ in a variety of
models
since it depends mainly on the Coulomb interaction. In particular, we
will use the spherical Hartree-Fock model based upon the recent SkX set
of Skyrme parameters \cite{b6},
with the addition of charge-symmetry breaking
(CSB), SkX$_{csb}$ \cite{b7}.
With the addition of CSB these calculations are able
to reproduce the measured displacement energies for all but the
lightest nuclei to within an rms
deviation of about 100 keV \cite{b7}. In the $  A=41-75  $
mass region the mass
(binding energy) of most of the neutron-rich nuclei are experimentally
usually known to within 100 keV or better (the only exception
being $^{71}$Br for which we use the AWE).
Thus we combine the
experimental binding energy for the neutron-rich nucleus
$  BE(A,T_{z}^{>})_{exp}  $ together with the Hartree-Fock
value for $  D(A,T)_{HF}  $ to provide an extrapolation
for the proton-rich binding
energy:

\begin{equation}
BE(A,T_{z}^{<}) = D(A,T)_{HF} + BE(A,T_{z}^{>})_{exp}.
\end{equation}

\noindent
  The method is similar to the one used by Ormand \cite{b4} for the
proton-rich nuclei with $  A=46-70  $. In \cite{b4} the displacement
energies are based upon shell-model configuration mixing which
includes Coulomb and CSB interactions with parameters for the
single-particle energies and strengths which are fitted to this mass
region. In the present work, which covers the region $  A=41-75  $,
the displacement energies are based upon
Skyrme Hartree-Fock calculations with a global
set of parameters which are determined from the properties of
closed-shell nuclei and nuclear matter. The CSB part of the interaction
has one parameter which was adjusted to reproduce the displacement
energies in the $  A=48  $ mass region \cite{b7}.

The displacement energies for all but the lightest nuclei
can be reproduced with the constant CSB interaction given in
\cite{b7}, and we use the same CSB interaction for the
extrapolations to higher mass discussed here.

The calculations presented here are relevant for the masses of
proton-rich nuclei via their connection with their mirror neutron-rich
analogues. We are not able to improve upon the masses of nuclei with
$N=Z$, and as will be discussed, the relative large errors which remain
for the $^{64}$Ge and $^{68}$Se masses provide now the dominant uncertainty
in the rp-process calculations.

Details of the Hartree-Fock calculations will be discussed, and a
comparison between the calculated and experimental displacement energies
for the $  A=41-75  $ mass region will be made. Then the extrapolations
for
the proton-rich masses and the associated one- and two-proton separation
energies will be presented. The proton drip line which is established by
this extrapolation will be compared to experiment, and the nuclei which
will be candidates for one- and two-proton decay will be discussed.
Finally we explore the significance of the new extrapolation for the rp process
in type I X-ray bursts.

\section{Displacement-Energy Calculations}

The SkX$_{csb}$ interaction is used to carry out Hartree-Fock
calculations
for all nuclei in the range $  Z=20-38  $ and $  N=20-38  $. The binding
energies are then combined in pairs to obtain theoretical displacement
energies for $  A=41-75  $ and $  T=1/2  $ to $  T=4  $:

\begin{equation}
D(A,T)_{HF} = BE(A,T_{z}^{<})_{HF} - BE(A,T_{z}^{>})_{HF}.
\end{equation}

The calculation is similar to those presented in \cite{b7}, but several
refinements are made. The single-particle states in proton-rich nuclei
become unbound beyond the proton-drip line. In the nuclei we
consider they are unbound by up to about 2
MeV. Since 2 MeV is small compared to the height of the Coulomb barrier
(about 6 MeV at a radius of 7 fm), the states are ``quasi-bound" and
have a
small proton-decay width (on the order of keV or smaller). To obtain the
quasi-bound wave functions we put the HF potential in a box with a
radius of 20 fm and a depth of 20 MeV. In all cases we consider, the
dependence of the results on the form of the external potential is
negligible as long as the radius is greater than about 10 fm and the
potential depth is greater than about 10 MeV.

In \cite{b7} the occupation numbers of the spherical valence
states were filled sequentially, and in this mass region they
always occur in the order f$_{7/2}$, p$_{3/2}$, f$_{5/2}$, p$_{1/2}$
and g$_{9/2}$. We have improved on this scheme by
carrying out an exact pairing (EP) calculation \cite{b8}
at each stage of the
HF iteration. The exact pairing model has recently been discussed in
\cite{b8}. The EP method
uses the single-particle energies from the HF calculation
together with a fixed set of $  J=0, T=1  $ two-body matrix elements and
gives the orbit occupations and the pairing correlation energy.
The orbit occupations are then used together with the HF radial
wave functions to calculate the nucleon
densities which go into the Skyrme energy density functional.
This procedure is iterated until convergence (about 60 iterations).
The pairing is calculated for protons and neutrons with the same
set of two-body matrix elements taken from the FPD6 interaction
for the pf shell \cite{b9}
and the Bonn-C renormalized G matrix for the matrix elements
involving the g$_{9/2}$ orbit \cite{b10}.
For those nuclei we consider here,
the occupation of the g$_{9/2}$ orbit is always small. It is
known that deformed components of the 2s-1d-0g shell are essential
for the nuclear ground states above $  A=76  $
as indicated by the sudden drop in the energy of the 2$^{ + }$
state from 709 keV in $^{72}$Kr to 261 keV in $^{76}$Sr \cite{b11}.
Thus we do not
go higher than $  A=76  $. In addition, one cannot always use Eq.\ (2)
above $  A=76  $ since many of the masses of the neutron-rich nuclei
are not known experimentally.

The results we obtain are not
very sensitive to the strength of the pairing interaction
and the associated distribution of the nucleons
between the $  p  $ and $  f  $ orbits, since these orbits
have similar
rms radii and single-particle Coulomb shifts. For example,
a 20 percent change in the strength of the pairing interaction results
in displacement energy changes of less than 20 keV. If pairing
is removed, the displacement energies can change by up to
about 100 keV. Thus, at the level of 100 keV accuracy pairing
should be included, but it is not a crucial part of the 
model.

A final refinement has been to add
a Coulomb pairing contribution to the proton-proton
$  J=0  $ matrix elements. The two-body Coulomb matrix elements were
calculated in a harmonic-oscillator basis. The Coulomb pairing
is then defined as the difference of the diagonal $  J=0  $ matrix
elements
from the $  (2J+1)  $ weighted average (which corresponds to the
spherical part of the Coulomb potential which is in the HF part
of the calculation). The Coulomb pairing matrix elements are
50-100 keV.

\begin{figure}[ht]
\includegraphics[]{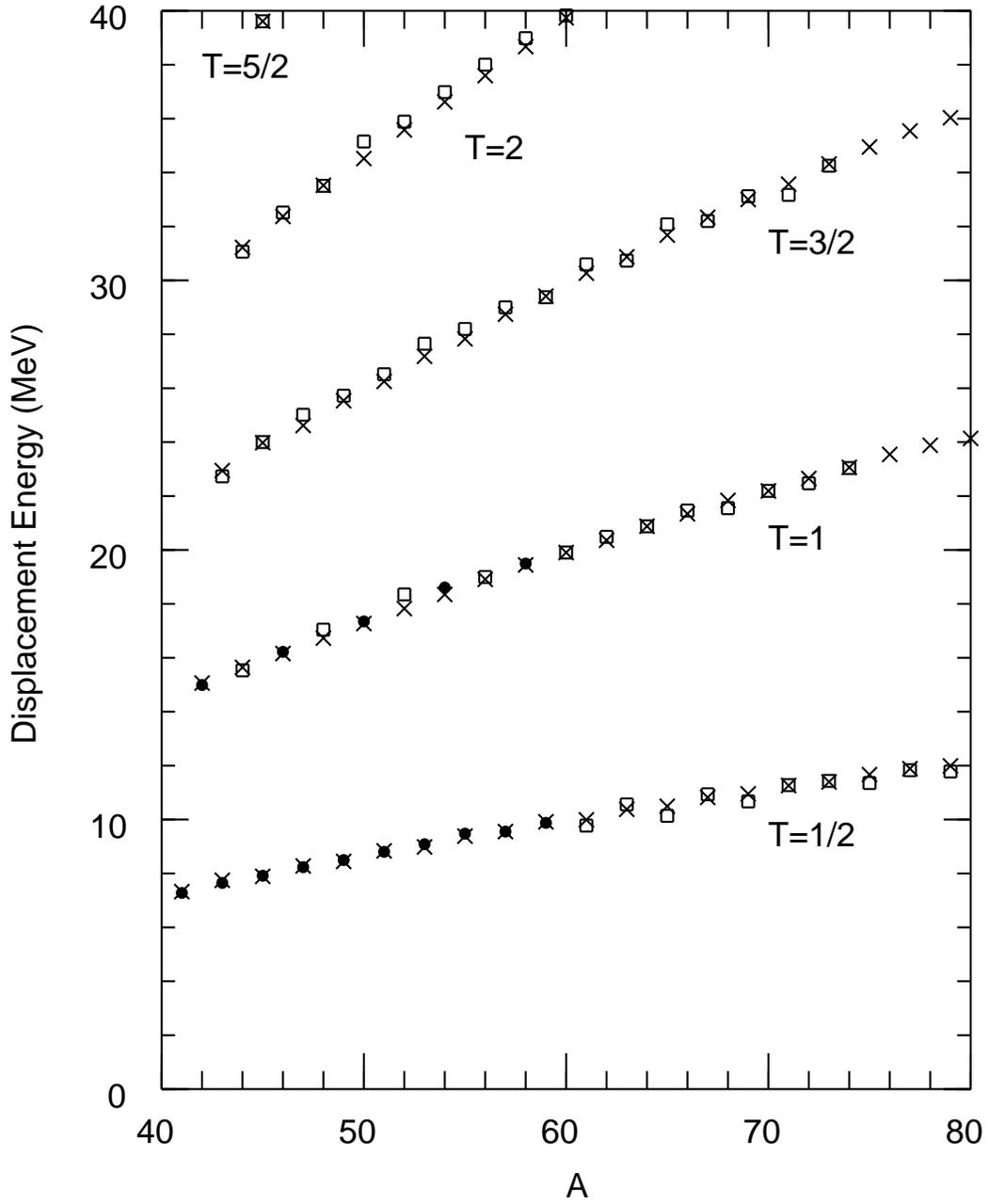}
\caption{Calculated displacement energies (crosses) as a function of mass
number.
They are compared to experimental data (filled circles) and to values
based upon the Audi-Wapstra extrapolations (squares).
\label{f1}}
\end{figure}

\begin{figure}[ht]
\includegraphics[]{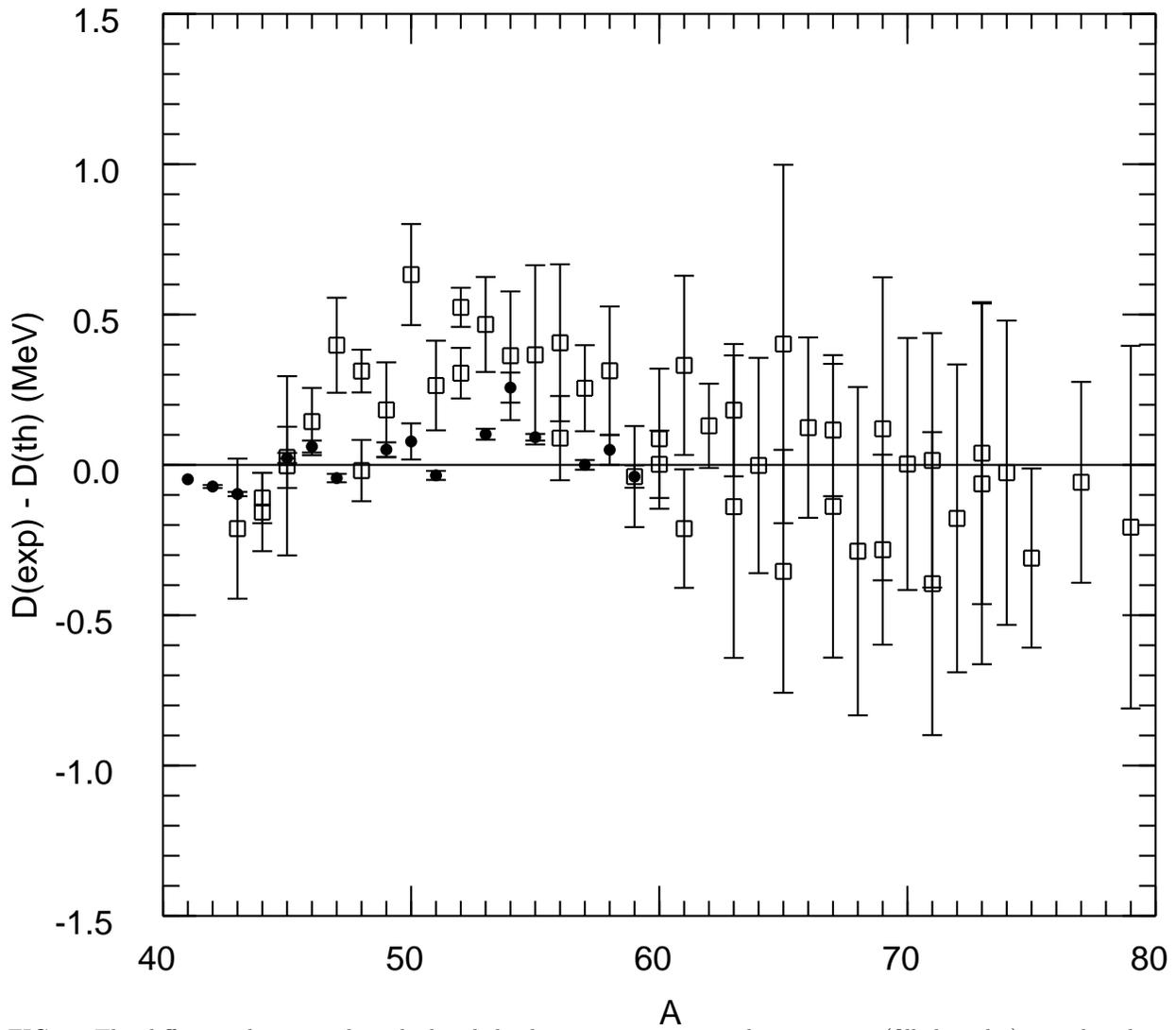}
\caption{The difference between the calculated displacement energies
and experiment (filled circles) or values based on the
the Audi-Wapstra extrapolations (squares).
\label{f2}}
\end{figure}

In Fig.~\ref{f1} the calculated displacement energies (crosses) are shown
in comparison with experiment (filled circles) in cases where
both proton- and neutron-rich masses have been measured and with the
AWE (squares) in
cases where the mass of the proton-rich nucleus is based upon the
AWE. The corresponding differences between
experiment and theory are shown in Fig.~\ref{f2} including the experimental
or
AWE error bars. It can be seen that when the displacement
energy is measured the agreement with the calculation is excellent to
within an rms deviation of about 100 keV. The most exceptional deviation is
that
for $  A=54  $ involving the $^{54}$Ni-$^{54}$Fe mirror pair;
a confirmation of the
experimental mass for $^{54}$Ni (which has a 50 keV error)
would be worthwhile. The comparison based
upon the AWE (squares) shows a much larger
deviation with typically up to 500 keV differences, but the AWE
error assumed is sometimes (but not always)
large enough to account for the spread.
The
implication of this comparison is that the error in the HF extrapolation
of the displacement energies is probably much less than the error in the
AWE of the displacement energies. In particular,
one notices in Fig.~\ref{f1} in the region $  A=60-75  $ that the
displacement
energy based upon the AWE shows a small
oscillation which is not present in the HF calculation and which
is not present in the experimental data for $  A<60  $.

\section{Proton-Rich Masses and Separation Energies} \label{SecSp}

The next step is to use Eq.\ (2) to calculate the binding energy
of the proton-rich nuclei based upon the HF calculation of the
displacement energy together with the experimental binding energy
of the neutron-rich nucleus \cite{b1,TBZ01}. The only neutron-rich nucleus
whose mass is not
yet experimentally measured is $^{71}$Br for which we use the AWE value.
The binding energies for the HF extrapolations for the
proton-rich nuclei are given an
error based upon the experimental error of the neutron-rich binding
energy folded in quadrature with an assumed theoretical error
of 100 keV.

The HF extrapolated set of binding energies for
proton-rich nuclei together with the experimental binding
energies for nuclei with $  N=Z  $ and neutron-rich nuclei provides
a complete set of values from which the one- and two-proton
separation energies are obtained. The masses for the
$  N=Z  $ nuclei $^{66}$As,
$^{68}$Se,
$^{70}$Br,
 are not measured and we use the AWE value.
The mass for $^{74}$Rb has a relatively large experimental error.

\begin{figure}[ht]
\includegraphics[height=7in]{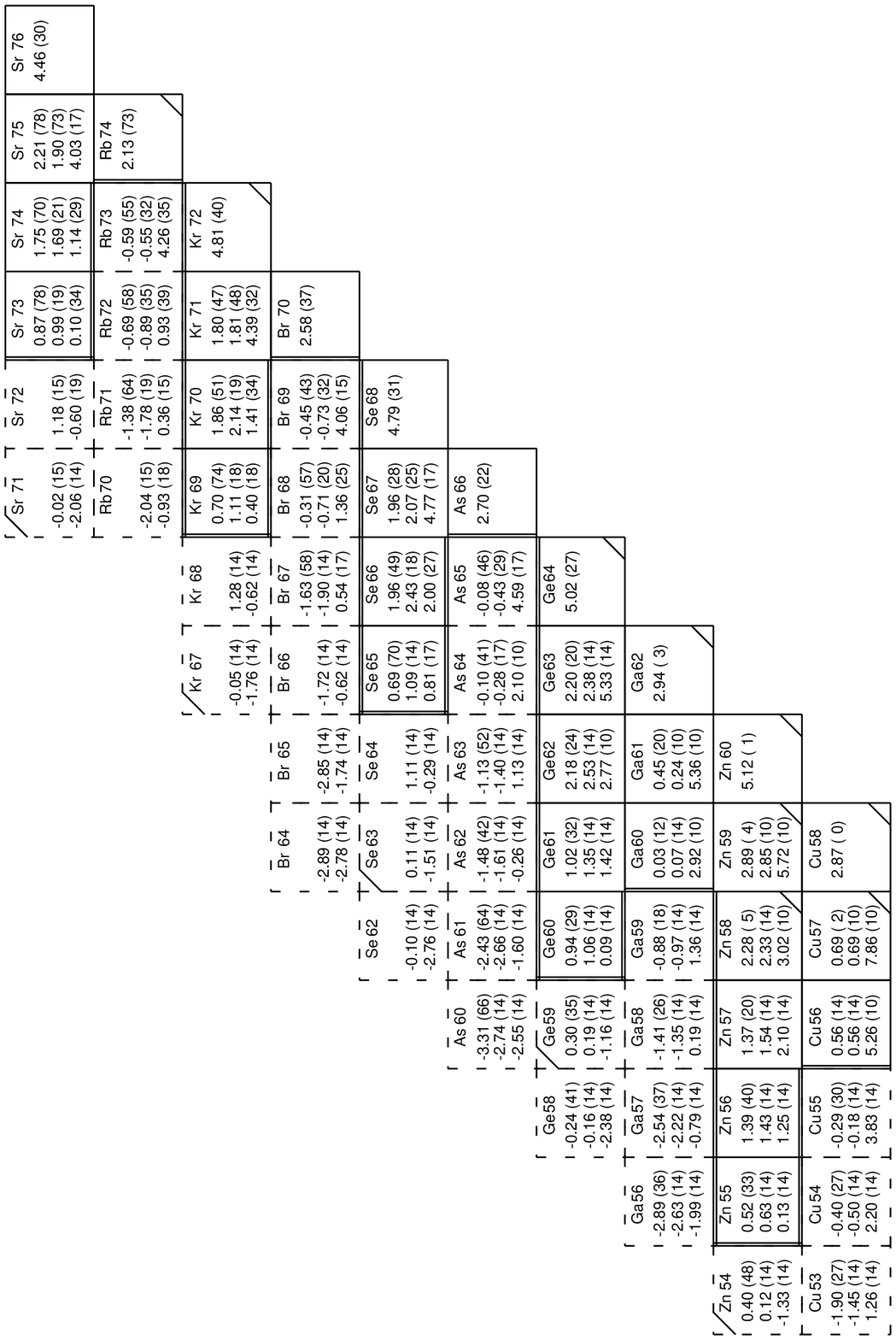}
\caption{A section of the mass chart for $  N=Z  $ and proton-rich nuclei
showing:
(line 1): the one-proton separation energy (followed by
the associated error) based upon AWE; (line 2):
the one-proton separation energy based upon the present HF calculations;
and (line 3): the two-proton separation energy based upon the HF
calculations. The line in the lower right-hand corner indicates that
the mass has been measured for this nucleus. A line in the upper
left-hand corner indicates that this nucleus is a candidate for diproton
decay.
\label{f3}}
\end{figure}

Results for the one- and two-proton separation energies are shown in
Fig.~\ref{f3}. The first line in each box is the one-proton separation
energy
(and the associated error)
based upon the AWE with the associated error. The second line is the
one-proton separation energy based upon the HF extrapolation, and the
third line is the two-proton separation energy based upon the HF
extrapolation. The error in the separation energies is the error for the
binding energies of the parent and daughter nuclei folded in quadrature.

The double line in Fig.~\ref{f3} is the proton-drip line beyond which the
one-proton separation energy and/or the two-proton separation energy
becomes negative. However, due to the Coulomb barrier, some of the
nuclei beyond the proton-drip line may have lifetimes which are long
enough to be able to observe them in radioactive beam experiments.
The observation of $^{65}$As in the experiment of Blank et
al. \cite{b12} excludes half-lives which are much shorter than 1 $\mu$s
which indicates that it is unbound by less than 400 keV. The
identification of $^{65}$As as a $\beta$-emitter by Winger et al.
\cite{WBB93}
together with the non-observation of emitted protons by Robertson et al.
\cite{RRL90} indicates that it is unbound by less than 250~keV.
Both limits are
compatible (within error) with the HF results given in Fig.~\ref{f3}. The
non-observation of $^{69}$Br in the radioactive beam experiments of
Blank et al. \cite{b12} and Pfaff
et al. \cite{b13} means that its lifetime is less than 24 nsec which
implies that it is proton unbound by more than 500 keV \cite{b13}.
This is compatible with the HF result shown in Fig.~\ref{f3}. The
non-observation of $^{73}$Rb in the experiments of
Mohar et al. \cite{MBB91}, Jokinen et al. \cite{JOA96}, and Janas et al.
\cite{b14}
gives an upper limit of 30 nsec for the half-life which implies that
$^{73}$Rb is proton unbound by more than 570 keV, again in agreement
(within error) of the present HF result. Thus all of the current
experimental data are consistent with our calculations.

The proton-drip line has not yet been reached for most $  Z  $ values.
Beyond the
proton-drip line there are several candidates for nuclei
which should be explored for one-proton emission: $^{54}$Cu, $^{58}$Ge,
$^{64}$As, $^{68}$Br, $^{69}$Br, $^{72}$Rb and $^{73}$Rb.
The most
promising candidates for the illusive diproton emission
(in addition to $^{48}$Ni \cite{b2,b4}) are $^{64}$Zn,
$^{59}$Ge,
$^{63}$Se, $^{67}$Kr and $^{71}$Sr. Estimated lifetime ranges for these
diproton decays are given by Ormand \cite{b4}.

\section{Implications for the rp process}

The rp process beyond Ni plays a critical role during hydrogen burning
at high temperatures and densities on the surface of
accreting neutron stars in X-ray bursters and
X-ray pulsars \cite{WGI94,SAG97,SBC98,KHA99,SAB01}.
Nuclear masses are among the most important input
parameters in rp-process calculations, as they
sensitively determine the balance between proton capture and the inverse
process, ($\gamma$,p) photodisintegration. It is this ($\gamma$,p)
photodisintegration that prevents the rp process from continuing
via proton captures, once a nucleus close to the proton drip line is
reached.
This nucleus becomes then a "waiting point" as the rp process has to proceed
at least in part, via the slow $\beta^+$ decay. The effective lifetime of
the waiting points in the rp process determines
the overall processing time scale, energy generation, and the final
abundance distribution.
At a waiting point nucleus (Z,N), a local (p,$\gamma$)-($\gamma$,p)
equilibrium is established with the following isotones
(Z+1,N), (Z+2,N).
The effective proton capture flow destroying waiting point nuclei
and reducing their lifetime is then governed by the Saha equation and the
rate of the reaction leading out of the equilibrium. Because of the
odd-even structure of the proton drip line 2 cases have to be distinguished
\cite{SAG97}. For temperatures below $\approx$ 1.4~GK equilibrium is
only established with the following isotone (Z+1,N). In this case, the
destruction
rate of the waiting point nucleus via proton captures
$\lambda_{(Z,N)(p,\gamma)}$
is determined by
the Saha equation and the
proton capture rate on the following isotone (Z+1,N).
The total destruction rate of the waiting point nucleus (Z,N)
is then
given by the sum of proton capture and $\beta$-decay rates:
\begin{equation} \label{Eq2p1}
\lambda = \lambda_{\beta} +
Y_p^2 \rho^2 N_A^2
\left( \frac{2\pi \hbar^2}{\mu_{(Z,N)}kT} \right) ^{3/2}
\frac{G_{(Z+1,N)}(T)}{(2J_{\rm p}+1)G_{(Z,N)}(T)}
\exp{\left( \frac{Q_{(Z,N)(p,\gamma)}}{kT} \right)}
<\sigma v>_{(Z+1,N)(p,\gamma)}
\end{equation}
$\lambda_{\beta}$ is the $\beta$-decay rate
of nucleus (Z,N), $Y_p$ the
hydrogen abundance, $\rho$ the mass density, $J_p$ the proton spin,
$G_{(Z,N)}$ the partition
function of nucleus (Z,N), $T$ the temperature, $\mu_{(Z,N)}$ the reduced
mass of nucleus (Z,N) plus proton, $Q_{(Z,N)(p,\gamma)}$
the proton capture Q-value of the waiting point nucleus, and
$<\sigma v>_{(Z+1,N)(p,\gamma)}$ the proton capture rate on the
nucleus (Z+1,N).
For higher temperatures local equilibrium is maintained
between the waiting point nucleus (Z,N) and the next two following isotones
(Z+1,N) and (Z+2,N).
In this case, $\lambda_{(Z,N)(p,\gamma)}$ is given by the Saha equation and
the $\beta$-decay rate of the final nucleus $\lambda_{(Z+2,N) \beta}$,
and
the total destruction rate $\lambda$ of the waiting point nucleus becomes:
\begin{equation} \label{Eq2p2}
\lambda = \lambda_{\beta}+ Y_p^2 \rho^2 N_A^2
\left( \frac{2\pi \hbar^2}{kT} \right) ^{3}
\mu_{(Z,N)}^{-3/2} \mu_{(Z+1,N)}^{-3/2}
\frac{G_{(Z+2,N)}(T)}{(2J_{\rm p}+1)^2 G_{(Z,N)}(T)}
\exp{\left( \frac{Q_{(Z,N)(2p,\gamma)}}{kT} \right)}
\lambda_{(Z+2,N) \beta}
\end{equation}
In both cases, the destruction rate of a waiting point nucleus depends
exponentially either on its one-proton
capture Q-value $Q_{(Z,N)(p,\gamma)}$
or two-proton capture
Q-value $Q_{(Z,N)(2p,\gamma)}$.
Nuclear masses therefore play a critical role in determining
the rp-process waiting points and their
effective lifetimes.

It has been shown before
that the most critical waiting point nuclei for the rp process beyond Ni
are $^{64}$Ge, $^{68}$Se and $^{72}$Kr \cite{SAG97}.
With the exception of $^{56}$Ni and $^{60}$Zn,
these nuclei are by far the longest-lived isotopes in the rp-process path.
The reason for those three nuclei being the most critical ones is that
with increasing charge number the $N=Z$ line moves closer to the proton
drip line and away from stability. Therefore, proton capture Q-values
on even-even $N=Z$ nuclei, which are favored in the rp process because of
the odd-even structure of the proton drip line, decrease with increasing
charge number, while the $\beta$-decay Q-values become larger. $^{64}$Ge,
$^{68}$Se and $^{72}$Kr happen to be located in the "middle", where
proton capture Q-values are already low enough to suppress proton captures
and
allow $\beta$ decay to compete,
but at the same time $\beta$-decay Q-values are still
small enough for half-lives to be long compared to rp-process time scales.
The critical question is to what degree proton captures can reduce the long
$\beta$-decay lifetimes of $^{64}$Ge (63.7~s half-life),
$^{68}$Se (35.5~s half-life) and $^{72}$Kr (17.2~s half-life).
As Eqs.~\ref{Eq2p1} and \ref{Eq2p2} show,
the answer depends mainly on
the one- and two-proton capture Q-values. Unfortunately, experimental
data exist for none of the relevant Q-values. The only available
experimental information are upper limits of the one-proton capture
Q-values of $^{68}$Se and $^{72}$Kr from the non-observation of
$^{69}$Br \cite{b12,b13} and $^{73}$Rb \cite{MBB91,JOA96,b14}, and
the lower limits on the one-proton capture Q-value on $^{65}$As from its
identification as a $\beta$-emitter  in radioactive
beam experiments (see Sec.~\ref{SecSp}).
While these data provide some constraints,
accurate Q-values are needed for the calculations and have to be predicted
by theory. The new masses
calculated in this work cover exactly this critical mass range, and
provide improved predictions for all the relevant Q-values in the
$A=64$--72 mass region (see Fig.~\ref{f3}).
As discussed in Sec.~\ref{SecSp}, all of our new predictions
are compatible with the existing experimental limits.

To explore the impact of the new mass predictions on rp-process models,
we performed calculations with a 1-D, one zone X-ray burst model
\cite{Bil97,SAB01}.
Ignition conditions are based on a mass accretion rate of 0.1 times
the Eddington accretion rate,
an internal heat flux from the neutron star surface of
0.15~MeV/nucleon, an accreted matter
metallicity of 10$^{-3}$ and a neutron star
with 1.4 solar masses and 10~km radius.

In principle, proton separation energies can influence the reaction flow
in two ways. First, they affect the forward to reverse rate ratios for
proton capture reactions and
the local (p,$\gamma$)-($\gamma$,p) equilibria through the $\exp(Q/kT)$ term
in the Saha equation (in Eqs.~\ref{Eq2p1} and \ref{Eq2p2}). 
This leads to an exponential mass dependence of the waiting point lifetimes. 
Second, theoretical
predictions of reaction rates $<\sigma v>$ (in Eq.~\ref{Eq2p1})
depend also on the adopted Q-values. 
In this work we choose to take into account both effects.
To explore the impact of Q-value uncertainties on 
proton capture reaction rate calculations
we use the statistical model code SMOKER
\cite{SAG97}.
Even though the nuclei in question are close to the proton drip line
a statistical approach is justified in most cases 
because reaction rates tend to become important only for
larger Q-values when a local (p,$\gamma$)-($\gamma$,p) equilibrium
cannot be established. Then the
level density tends to be sufficient for the
statistical model approach. Based on the new reaction rates
we then use our new Q-values to recalculate 
($\gamma$,p) photodisintegration rates via detailed balance
as discussed in \cite{SAG97}.

\begin{figure}[ht]
\includegraphics[height=7in]{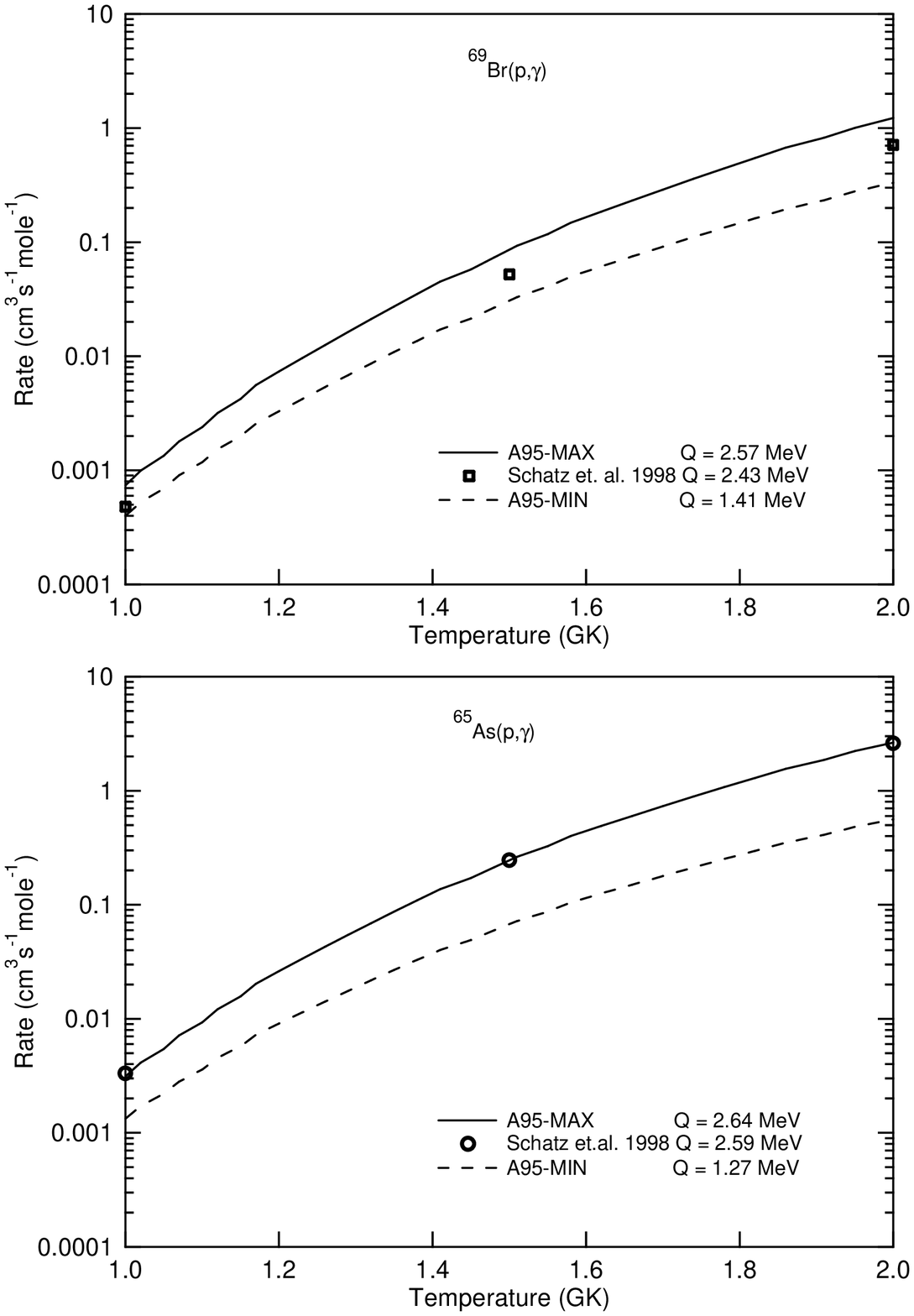}
\caption{The calculated $X(p,\gamma)$ rates for $X=^{65}$As and $X=^{69}$Br with associated Q-values shown in the legend. The astrophysical reaction rates were
calculated with the statistical model code SMOKER.
\label{f4}}
\end{figure}

For the relevant temperature range between 1-2~GK,
our new proton capture reaction rates vary in most cases
not more than a factor of two within the explored mass uncertainties.
An exception among the relevant reaction rates are the
proton capture rates on $^{65,66}$As, $^{69,70}$Br, and $^{73,74}$Rb.
These rates show a somewhat stronger variation of typically a factor of 4 -- 6
as the associated
proton capture Q-values are particularly uncertain. 
Fig.~\ref{f4} shows two examples for the Q-value dependence of 
statistical model reaction rates. Generally, a larger Q-value 
leads to larger rates, as the higher excitation energy of 
the compound nucleus opens up more possibilities for its decay.
For reference, Fig.~\ref{f4} also shows the rates listed in
\cite{SAG97}, which had been calculated using Q-values from 
the Finite Range Droplet Mass model (FRDM1992) \cite{MNM95}.

To disentangle the different effects of mass uncertainties
quantitatively we performed test calculations
in which changes in masses were only
taken into account in the calculation of the ($\gamma$,p) photodisintegration
rates, while the proton capture rates were kept the same. These test
calculations lead to very similar
luminosity and burst time scale variations as presented in this paper.
Discrepancies were at most
8\% in the luminosity and 0.1\% in the burst timescale.
This can be understood from Eq.~\ref{Eq2p1} and \ref{Eq2p2}.
For example, a change of 1.37~MeV
in the proton capture Q-value changes the $^{65}$As reaction rate 
and therefore the lifetime of the $^{64}$Ge waiting point nucleus
by a
factor of 3--4 (see Fig.~\ref{f4} and Eq.~\ref{Eq2p1}).
However, the same
1.37~MeV Q-value change in the  $\exp(Q/kT)$ term in Eq.~\ref{Eq2p2}
would result in a lifetime change of 6 orders of magnitude 
(for a typical $kT=100$~keV).
We therefore
conclude that the impact of mass uncertainties on rp process
calculations through changes in theoretical reaction rate calculations
within the statistical model is much smaller than the 
impact through changes in (p,$\gamma$)/($\gamma$,p) reaction rate ratios.

\begin{figure}[ht]
\includegraphics[height=7in]{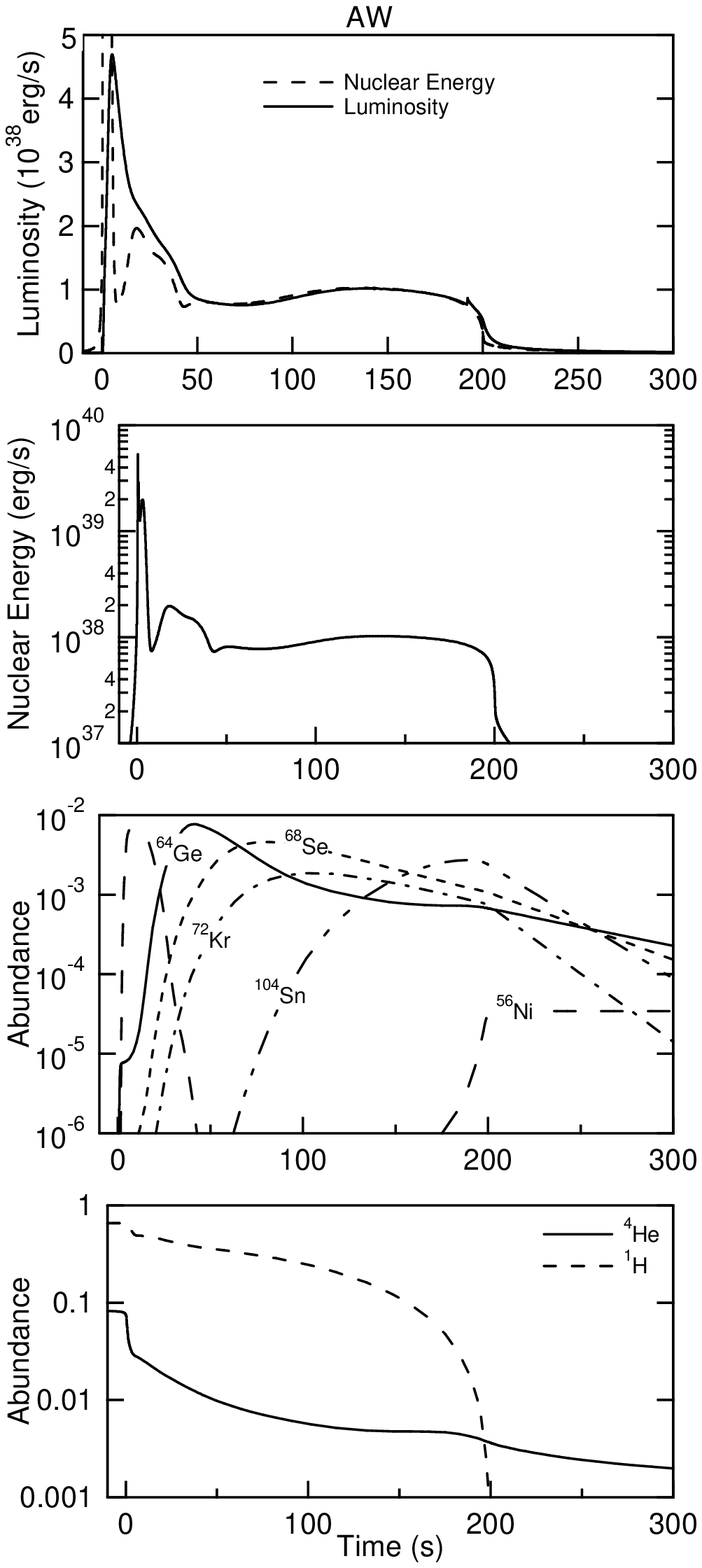}
\includegraphics[height=7in]{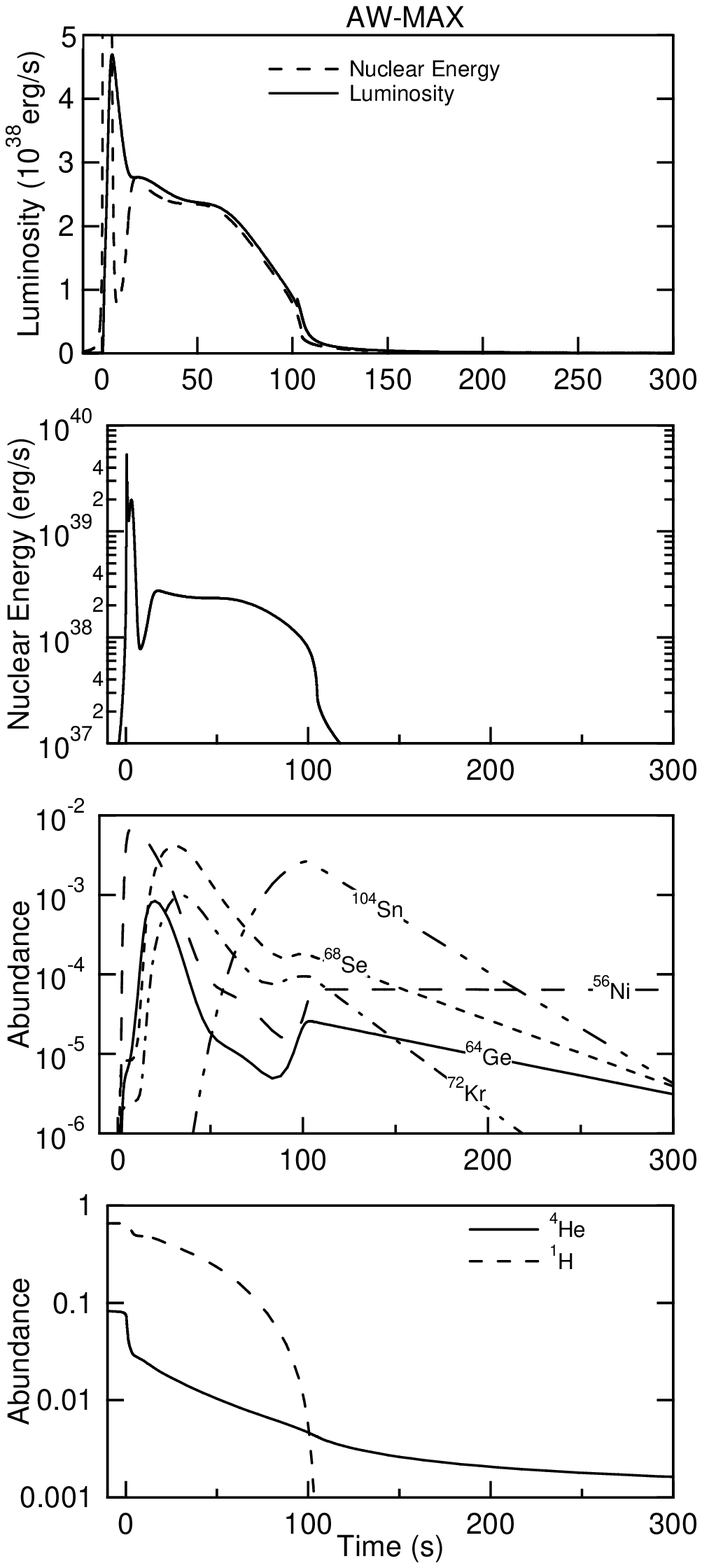}
\caption{ Luminosity, nuclear energy generation rate, and the abundances of
hydrogen, helium, and the critical waiting point nuclei
as functions of time as predicted by our X-ray burst model for different 
sets of proton capture Q-values. Shown are results for the sets SkX-MIN, SkX, and
SkX-MAX for the smallest, the recommended, and the largest proton capture 
Q-values within the error bars of the mass predictions of this work. 
A corresponding series is shown for the Audi \& Wapstra 1995 mass evaluation
(AW-MIN, AW, and AW-MAX). 
The $^{104}$Sn abundance indicates the operation of the SnSbTe cycle. 
Also, for comparison, the nuclear energy generation rate is shown as a dashed
line together with luminosity, though it is off the scale shown during the peak
of the burst.  The mass of the accreted layer is 5.0$\cdot 10^{21}$~g.
\label{f5}}
\end{figure}

\begin{figure}[ht]
\includegraphics[]{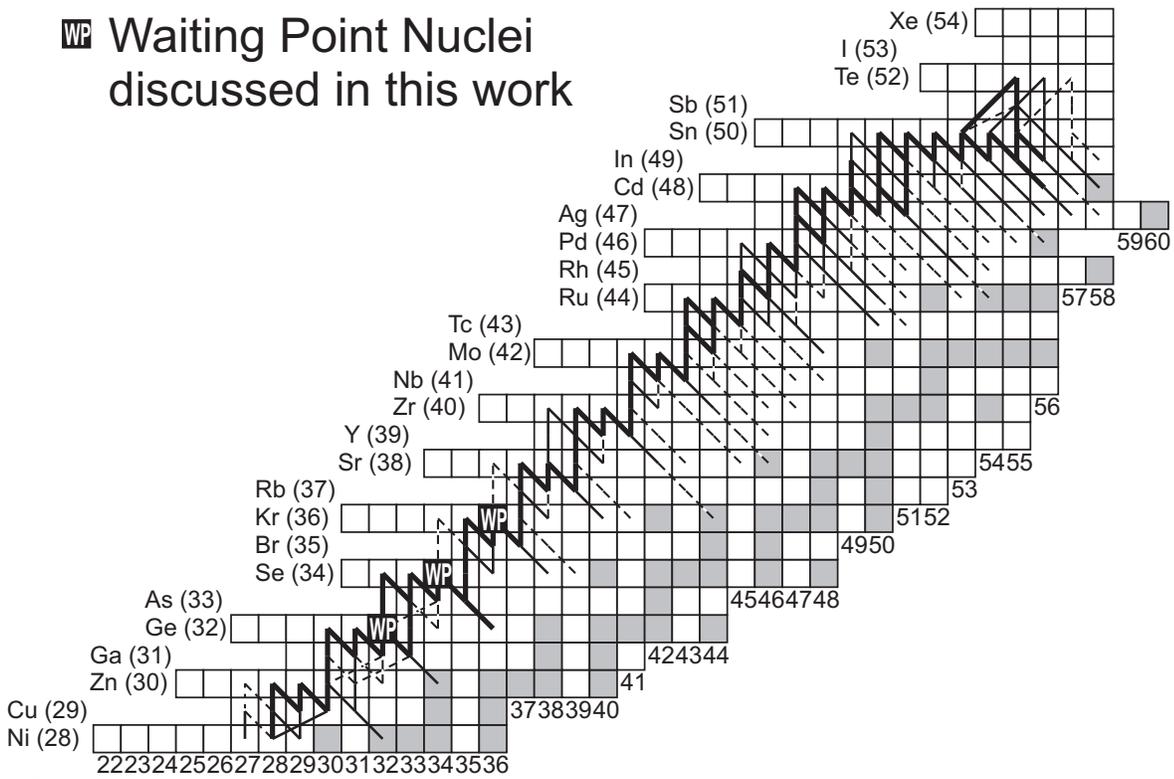}
\caption{The time integrated reaction flow beyond Ni during an X-ray burst
calculated
on the basis of our new mass predictions. Shown are flows of more than 10\%
(thick solid line), 1\%-10\% (thin solid line), and 0.1\%-10\% (dashed line)
of the flow through the 3$\alpha$ reaction. The key waiting points discussed
in this work are marked as well.
\label{f6}}
\end{figure}

The following calculations
were performed with different assumptions on masses beyond the $N=Z$ line
from $Z=30$-38: SkX based on the mass predictions of this work, SkX-MIN with
all proton capture Q-values set to the lowest value,
and SkX-MAX with all proton capture Q-values set to the highest values
within the error bars of our binding energy predictions. 
A similar set of calculations has been performed for the
mass extrapolations of Audi and Wapstra
1995 \cite{b1} (AW95) and are labeled AW, AW-MIN, and AW-MAX.
Fig.~\ref{f5}
shows the X-ray burst light curve, the nuclear energy generation rate, the
abundances of the most important waiting point nuclei and the hydrogen and
helium
abundances as a function of time for all our calculations.
As an example, Fig.~\ref{f6} shows the time integrated reaction flow
corresponding to the SkX calculation.
While the
$\alpha$p and rp processes below $^{56}$Ni are responsible for the rapid
luminosity rise at the beginning of the burst, processing through the slow
waiting points $^{64}$Ge, $^{68}$Se, $^{72}$Kr and
the operation of the SnSbTe cycle (indicated by the $^{104}$Sn abundance)
lead to an extended burst tail.  The rp process from $^{56}$Ni to $^{64}$Ge,
and the slowdown at $^{64}$Ge lead
to a pronounced peak in the energy generation rate around 50~s after
burst maximum. In principle the other
waiting points have a similar effect, but the corresponding peaks in the
energy production are much wider and therefore not noticeable.

Fig.~\ref{f7} compares X-ray burst light curves for
different assumptions on nuclear masses.
Generally, lower proton
capture Q-values enhance photodisintegration and favor the waiting point
nuclei
in local equilibria. Both effects lead to a slower reaction flow and
therefore
to less luminous but longer lasting burst tails.
Even though the uncertainties in our new mass predictions
are significantly smaller than in AW95,
they still allow for a burst length variation from $150$ -- $250~s$ and
a luminosity variation of about a factor of 2 (SkX-MIN and SkX-MAX).
The lower limit Q-value calculation with AW95 masses (AW-MIN)
is similar to our lower limit (SkX-MIN), but
the larger uncertainties in the AW95 masses
lead to large differences in the upper limits (SkX-MAX and AW-MAX)
and would imply significantly
shorter bursts with much more luminous tails (AW-MAX).
However, some of the large proton capture
Q-values in AW-MAX and to a lesser degree in SkX-MAX
are already constrained by the experiments on
$^{69}$Br and $^{73}$Rb. If those constraints are taken into account one
obtains
the AW-MAXEXP and SkX-MAXEXP calculations respectively, which are also
shown in Fig.~\ref{f7}. The SkX-MAXEXP and AW-MAXEXP light curves are
very similar.

The dependence of the light curves on the choice of proton capture Q-values
can be understood entirely from the changes in $\beta$-decay and
proton-capture
branchings of the main waiting points
$^{64}$Ge, $^{68}$Se, and $^{72}$Kr shown in Table ~\ref{TabBranch}.
The calculations
with the lower limits on proton capture Q-values (SkX-MIN and AW-MIN)
do not differ much as they
all predict that proton captures do not play a role. However, for the
upper limits sizable proton capture branches occur and lead to significant
reductions in the lifetimes of the waiting points. In our upper limit
(SkX-MAX)
we obtain
26\% proton capture on $^{68}$Se (via 2p capture) and 86\% proton capture
on $^{64}$Ge, while proton captures on $^{72}$Kr, with 8\%, play only a
minor role. These branchings become even larger for the AW95 upper limit
calculations
(AW-MAXEXP and AW-MAX).
Note that $\beta$ decay of $^{60}$Zn is
negligible (See Table ~\ref{TabBranch}) because proton capture
dominates for the whole range of nuclear masses
considered here.

\begin{figure}[ht]
\includegraphics[]{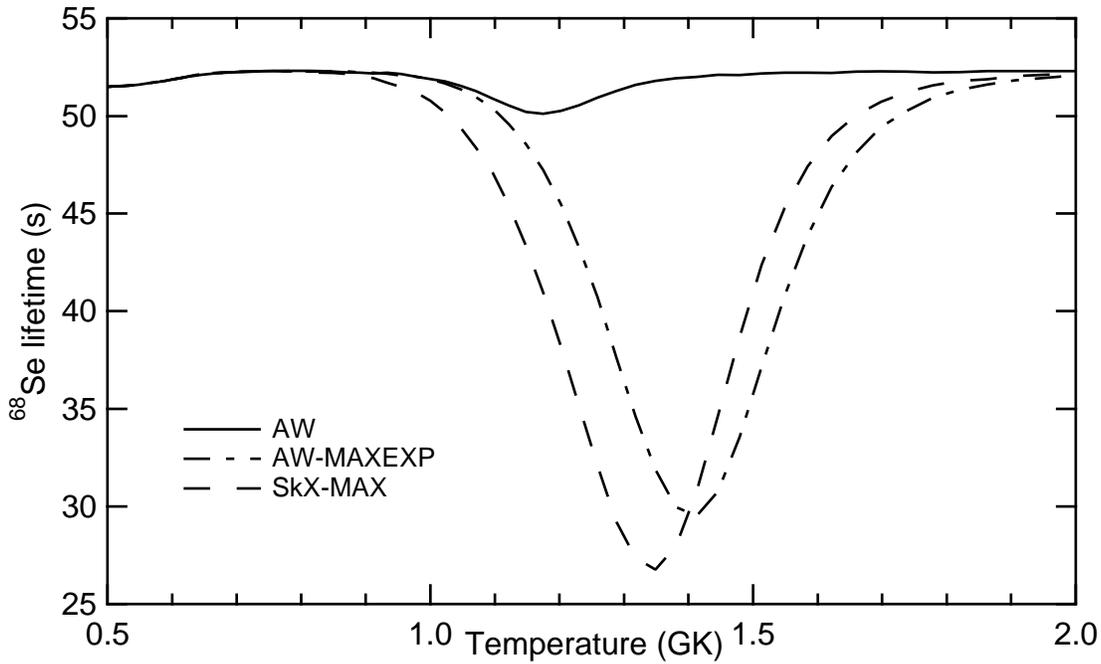}
\caption{The lifetime of $^{68}$Se against $\beta$-decay and
proton-capture for typical rp-process conditions during the burst tail
(hydrogen abundance 0.35, density 6$\cdot 10^{5}$~g/cm$^{3}$) for three
different
assumptions on proton capture Q-values on $^{68}$Se and $^{69}$Br:
Audi \& Wapstra 1995 recommended masses (AW), the largest proton capture
Q-values within the AW error bars but with experimental
constraints on the $^{68}$Se(p,$\gamma$) Q-value (AW-MAXEXP), and the
largest
proton capture Q-values within the error bars of the predictions from this
work (SkX-MAX).
\label{f8}}
\end{figure}

The importance of the one-proton capture Q-values in the determination of
lifetimes for rp-process waiting points has been discussed extensively
before
\cite{SAG97}. This importance is clearly expressed by the large changes in
branching
ratios and light curves when experimental constraints (which only exist for
one-proton separation energies) are imposed on the AW-MAX calculations
leading to AW-MAXEXP
(Fig.~\ref{f7} and Table ~\ref{TabBranch}
). However, the two-proton capture Q-values
can be equally important.  For example, the proton capture branching on
$^{68}$Se
changes by an order of magnitude from 2\% in AW to 15\% in AW-MAXEXP. This
change
is entirely due to the change in the $^{70}$Kr proton separation energy
from 1.86~MeV in AW to 2.4~MeV in AW-MAXEXP as the
proton capture Q-value on $^{68}$Se is very similar (only 
0.05~MeV difference).
The reason for this sensitivity is 
the onset of photodisintegration of $^{70}$Kr that depends very sensitively on
its proton separation energy. As soon as temperatures are sufficiently high
for $^{70}$Kr($\gamma$,2p)$^{68}$Se to play a role, $^{68}$Se, $^{69}$Br,
and $^{70}$Kr are driven into a local (p,$\gamma$)-($\gamma$,p) equilibrium.
With rising temperature the
proton capture on
$^{68}$Se drops then quickly to zero, because the temperature independent
and slow $\beta$ decay of $^{70}$Kr in Eq.~\ref{Eq2p2} cannot provide
a substantial leakage out of the equilibrium. This is different from
the situation at lower temperatures described by Eq.~\ref{Eq2p1} where a
lower
equilibrium abundance of $^{69}$Br at higher temperatures can be somewhat
compensated
by the increasing proton capture rate on $^{69}$Br. This effect is
illustrated
in Fig.~\ref{f8} which shows the lifetime of $^{68}$Se against
proton
capture and $\beta$ decay as a function of temperature
for different choices of proton capture Q-values. The lifetime equals the
$\beta$-decay lifetime for low temperatures because of slow proton capture
reactions,
and at high temperatures because of the photodisintegration effect discussed
above. For the AW masses, the low proton separation energy of $^{70}$Kr
leads to strong photodisintegration already at temperatures around 1.15~GK
before proton captures can play a role. Therefore proton captures never
reduce
the lifetime significantly. For AW-MAXEXP, the only change is a
larger $^{70}$Kr proton separation energy of 2.4~MeV. Though $^{69}$Br is
unbound by 500~keV, proton captures can reduce the lifetime of $^{68}$Se
by about a factor of two around 1.4~GK before photodisintegration sets in
and starts inhibiting further proton captures. This can be compared with
the upper limits of our predictions for proton separation energies
(SkX-MAX).
The larger proton separation energy of $^{69}$Br allows an onset of
proton captures at slightly lower temperatures, but the lower proton
separation energy of $^{70}$Kr leads also to an onset of photodisintegration
at somewhat lower temperatures thus effectively shifting the drop in
lifetime by about 0.1~GK. Note that it is not only the amount of lifetime
reduction, but also how well necessary conditions
match the actual conditions during the cooling of the X-ray burst that
determine
the role of proton captures and therefore the overall time scale of the rp
process.
As Fig.~\ref{f8} shows, both depends sensitively on the nuclear
masses.

\begin{figure}[ht]
\includegraphics[]{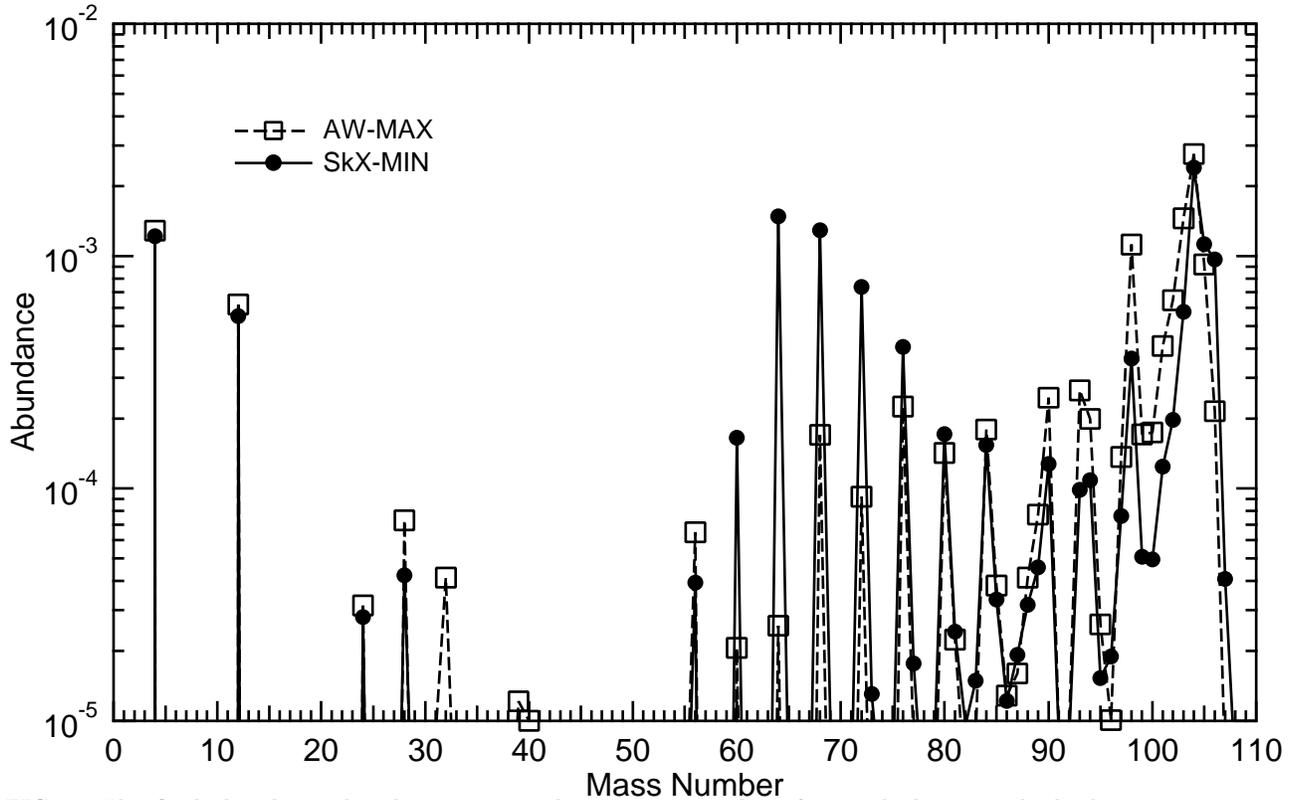}
\caption{The final abundance distribution summed over mass numbers for
a calculation with the lowest proton capture Q-values within the
uncertainties
of the mass predictions of this work (SkX-MIN) and with the largest
proton capture Q-values within the uncertainties of AW95 (AW-MAX).
\label{f9}}
\end{figure}

A long-standing question is how the nuclear physics, and in particular the
properties of the long-lived waiting points $^{64}$Ge, $^{68}$Se, and $^{72}$Kr affect
the end-point of the rp process.
Even for
our lowest proton capture Q-values where proton captures on
$^{68}$Se and $^{72}$Kr become negligible, we still find that the
rp process reaches the SnSbTe cycle \cite{SAB01}.
Fig.~\ref{f9} shows the final
abundance distribution for the two extreme cases -
our calculation with the slowest (SkX-MIN)
and the fastest (AW-MAX) reaction flow. In both cases, the most abundant
mass number
is $A=104$, which is due to accumulation of material in the SnSbTe cycle at
$^{104}$Sn. The main difference between the abundance patterns are the
abundances that directly relate to the waiting points at $A=64$, 68, and 72
and
scale roughly with the waiting point lifetime. In addition, for AW-MAX
nuclei in the $A=98$--103 mass range are about a factor of 3 more abundant
because of the faster processing
and the depletion of $A=64$, 68, and 72.

\section{Summary and Conclusions}

We have made a new set of predictions for the masses of
proton-rich nuclei on the basis of the displacement energies
obtained from spherical
Hartree-Fock calculations with the SkX$_{csb}$ Skyrme
interaction \cite{b6,b7}.
SkX$_{csb}$ provides a large improvement in
the displacement energies over those obtained with other Skyrme
interactions via the addition of a one-parameter charge-symmetry
breaking component \cite{b7}. 
A comparison with the experimental displacement
energies measured in the mass region A=41-59 indicates that the accuracy
of the calculated displacement energies is about 100 keV. We thus use
this as a measure of the uncertainty expected for the higher mass region
of interest in this paper. Experimental masses for some
proton-rich nuclei in the mass region A=60-70 will be required to test
our predictions. At the upper end, we may expect some deviation due
to the very deformed shapes which involve the excitation
of many $pf$-shell nucleons into the $g_{9/2}$ ($sdg$) shell
which go beyond our spherical approach.
In addition to the application to the rp-process, we have
discussed the implication of the present model for the proton
drip line. The most
promising candidates for diproton emission
are $^{64}$Zn,
$^{59}$Ge,
$^{63}$Se, $^{67}$Kr and $^{71}$Sr.

Our rp-process calculations based upon the masses obtained in the
present model and those obtained from the 
Audi-Waptra mass extrapolations demonstrate clearly the sensitivity of X-ray burst tails
on nuclear masses at and beyond the $N=Z$ line between Ni and Sr.
Such a sensitivity on the Q-values
for proton capture on $^{64}$Ge and $^{68}$Se has been pointed out before
by Koike et al. \cite{KHA99} based on a similar X-ray burst model.
However, Koike et al. \cite{KHA99} used a limited reaction network including
only
nuclei up to Kr. As we show in this paper, this
is not sufficient for any assumption on nuclear masses, and as a consequence
we find very different light curves and final abundances.

\begin{figure}[ht]
\includegraphics[height=7in]{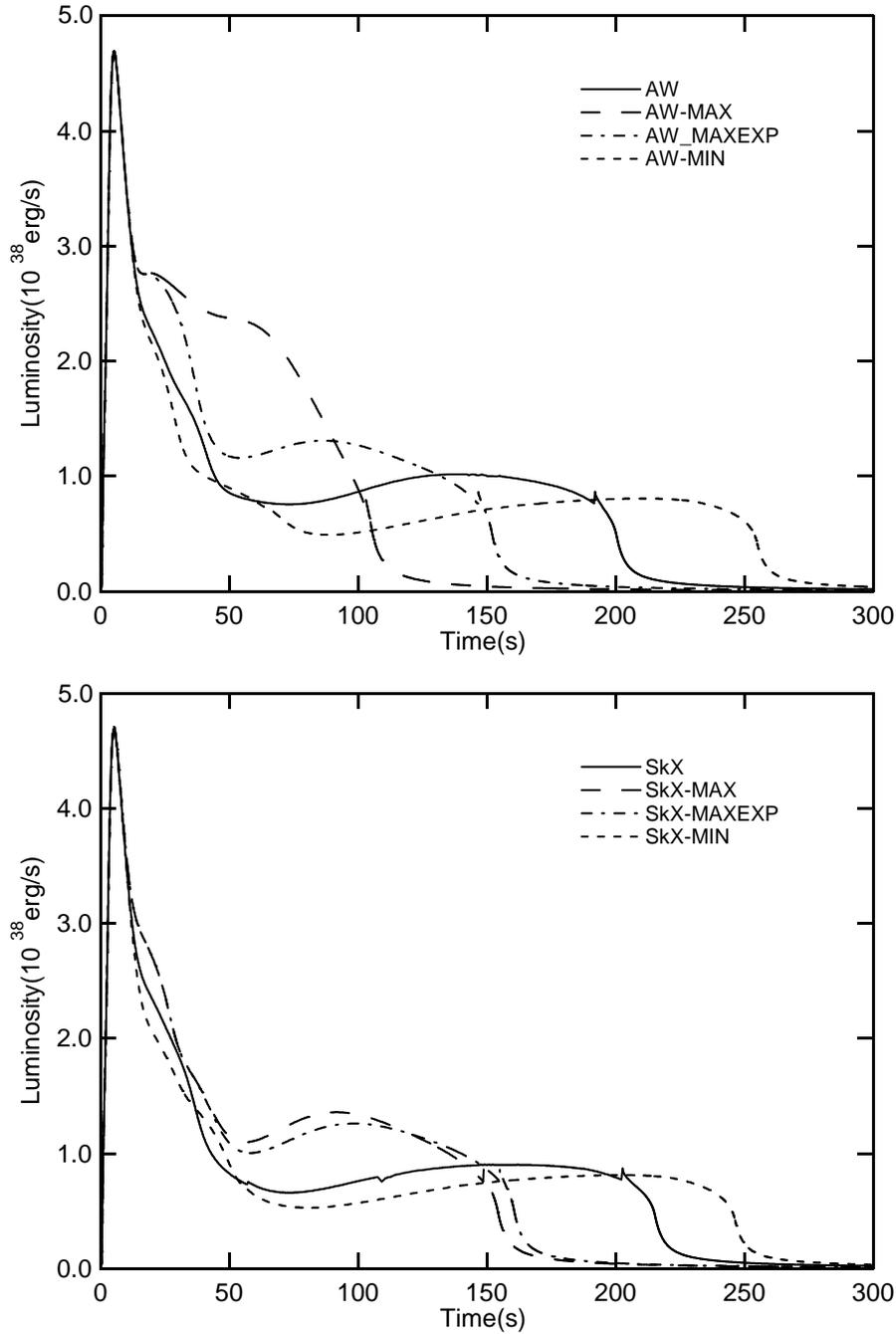}
\caption{X-ray burst luminosity as functions of time for model calculations
with different assumptions on proton capture Q-values in the Zn-Sr range:
results on the upper panel are
based on the Audi \& Wapstra 1995 recommended masses (AW) and the largest
(AW-MAX) and
smallest (AW-MIN) proton capture Q-values according to their error bars.
AW-MAXEXP
is identical to AW-MAX, but takes into account experimental limits on the
proton capture Q-values of $^{68}$Se and $^{72}$Kr. The lower panel shows the
same
set of calculations based on the mass predictions of this work (SkX...). The
mass of the accreted layer is 5.0$\cdot 10^{21}$~g.
\label{f7}}
\end{figure}

Our new calculation leads to tighter constraints 
on proton capture Q-values
as compared with the
AW95 mass extrapolations (see Fig.~\ref{f7}).
The first radioactive beam experiments including the
nonobservation of $^{69}$Br and $^{73}$Rb
have also begun to provide important constraints. If those experiments are
taken into account, our new predictions do not lead to substantially tighter
limits, with the exception of the proton capture on $^{64}$Ge, where
no experimental upper limit on the proton capture Q-value exists.
Our new calculations increase the minimum $\beta$ branching at
$^{64}$Ge by an
order of magnitude from 1\% to 14\%, leading to a lower limit
of the average $^{64}$Ge half-life in the rp process of 12.6~s instead of
0.9~s.
As a consequence, we predict a smooth and continuous drop in the light curve
during the first 30--40~s after the maximum, as opposed to the hump
predicted with AW-MAX.

However, uncertainties in the mass predictions are still too large to
sufficiently constrain the light curves and to determine the role that
proton captures play in the reduction of waiting point lifetimes. While
we find that within the errors of our mass predictions
proton capture on $^{72}$Kr is negligible,
our predicted average
proton capture branchings for $^{64}$Ge and $^{68}$Se still cover
a large range of 0.5\%--86\% and 0.0\% - 26\% respectively (of course this
is a model-dependent result - for example more hydrogen or a higher density
could strongly increase the proton capture branches). To a large extent
this is because of the large uncertainties in the masses of $N=Z$ nuclei
$^{64}$Ge (measured: 270~keV), $^{68}$Se (AW95 extrapolated: 310~keV),
and $^{72}$Kr (measured: 290~keV) \cite{b1} that cannot be determined
with the
method presented here. In addition, uncertainties in the masses of mirror
nuclei increase the errors for $^{73}$Rb (170~keV) and $^{70}$Kr (160~keV)
substantially beyond the $\approx$100~keV accuracy of our predicted Coulomb
shifts.
Overall, this results in typical uncertainties of the order of 300~keV for
several of the critical proton capture Q-values. 

To summarize, uncertainties in the masses of the nuclei that 
determine the proton capture branches on
$^{64}$Ge and $^{68}$Se represent a major nuclear physics uncertainty
in X-ray burst light curve calculations. 
The relevant nuclei are listed in the
upper part of Table~\ref{TabExp}
together with the currently available mass data and their uncertainties.
The proton capture branches on $^{60}$Zn
and $^{72}$Kr are of similar importance, but are sufficiently well
constrained by current experimental limits and theoretical calculations.
However, both the experimental and the theoretical limits are
strongly model dependent. Therefore, improved experimental mass data
would still be important to confirm the
present estimates. These nuclei are listed in the lower
part of Table~\ref{TabExp}. As discussed in Sec. \ref{SecSp}
there is experimental evidence 
indicating proton stability of all the nuclei listed, except for $^{69}$Br and 
$^{73}$Rb, which are probably proton unbound. Mass measurements of the proton bound
nuclei
could be performed with a variety of techniques including 
ion trap measurements, time of flight measurements, or $\beta$ decay 
studies. Recent developments in the production of radioactive beams 
allow many of the necessary experiments to be performed at existing
radioactive beam facilities such as ANL, GANIL, GSI, ISOLDE, ISAC, and the NSCL. 
Mass measurements of the proton unbound nuclei $^{69}$Br and $^{73}$Rb require their
population via transfer reactions from more stable nuclei, or
by $\beta$ decay from more unstable nuclei. Both are significantly 
more challanging as much higher beam intensities or the production of 
more exotic nuclei are required, respectively. 

Of course, burst timescales depend sensitively on the amount of hydrogen 
that is available at burst ignition.
The more hydrogen that is 
available the longer the rp process and the longer the burst tail 
timescale. In this work we use a model 
with a large initial hydrogen abundance (close to solar) to explore the impact
of mass uncertainties on X-ray burst light curves. 
This allows us
to draw conclusions on the uncertainties in predictions of the longest
burst timescales and the heaviest elements that can be produced in 
X-ray bursts. The former is important for example in light of recent observations
of very long thermonuclear X-ray bursts from GX 17+2 \cite{KHK01}, the latter
for the question of the origin of p nuclei discussed below. 
Nevertheless we expect a similar light curve sensitivity to masses for
other models as long as
there is enough hydrogen for the rp process to reach the $A=74-76$ mass
region.
In our one zone model we find that this requires about a 
0.35-0.45 hydrogen 
mass fraction at ignition. Even though the burst temperatures and densities
vary somewhat with the initial conditions we find shorter, but otherwise
very similar reaction 
paths governed by the same waiting point nuclei.
For bursts with initial hydrogen abundances below $\approx$ 0.3 the rp process does not 
reach the $A=60-72$ mass region anymore and the mass uncertainties discussed
in this work become irrelevant. 

Observed type I X-ray bursts show a wide variety of timescales ranging from 
10~s to hours. Our goal is to improve the underlying nuclear physics 
so that the observed burst timescales can be used to infer 
tight constraints on ignition conditions in type I X-ray bursts
such as the amount of hydrogen available for a given burst.
Such constraints
would be extremely useful as they could, for example, lead to constraints
on the impact of rotation and magnetic fields on the fuel distribution
on the neutron star surface as well as on the heat flux from the neutron
star surface \cite{Bil00,KHK00}. 
Our results indicate that without further
theoretical or experimental improvements on nuclear masses
it will not be possible to obtain such tight, quantitative constraints. 

Nevertheless, some qualitative conclusions can already be drawn on the basis
of our new mass predictions.
Our new results provide strong support for previous predictions that the rp
process
in the $A=64-72$ mass region slows down considerably leading to extended
burst tails \cite{SAB01}.
As a consequence, the long bursts observed 
for example
in GS~1826-24 \cite{KHK00}
can be explained by the presence of large amounts of hydrogen
at ignition and can therefore be interpreted as a 
signature of the rp process.

Even for
our lowest proton capture Q-values, when $^{68}$Se and $^{72}$Kr slow down
the rp process with their full $\beta$-decay lifetime
the rp process still reaches the SnSbTe cycle.
Clearly, such a slowdown of the rp process
does not lead
to a premature termination of the rp process as has been suggested
previously
(for example \cite{WGI94}), but rather extends the burst time scale
accordingly. 
As a consequence we find that hydrogen is completely consumed in 
our model.

However, a slower rp process will produce more nuclei in the $A=64$--72
range and
less nuclei in the $A=98$--103 mass range.
Interestingly,
among the most sensitive abundances beyond $A=72$
is $^{98}$Ru,
which is of special interest as it is one of the light p nuclei whose
origin in the universe
is still uncertain. p nuclei are proton rich, stable nuclei that cannot 
be synthesized by neutron capture processes. While standard p process
models can account for most of the p nuclei observed, they cannot produce
sufficient amounts of some light p nuclei such as $^{92,94}$Mo and $^{96,98}$Ru
(for example \cite{Ray95}).
Costa et al. \cite{Ray00} pointed out recently that a increase
in the $^{22}$Ne($\alpha$,n) reaction rate by a factor of 10-50 above the
presently recommended rate could help solve this problem, but recent 
experimental data seem to rule out this possibility \cite{Jae01}. 
Alternatively, X-ray bursts have been proposed as 
nucleosynthesis site for these nuclei \cite{SAG97,SAB01}.
An accurate determination of the $^{98}$Ru
production in X-ray bursts requires therefore
accurate masses in the $A=64-72$ mass range.
Further conclusions concerning X-ray bursts as a possible p process scenario
have to wait for future
self-consistent multi-zone calculations with the full reaction network, that
include the transfer of the ashes into the interstellar medium during
energetic
bursts.

Support for this work was provided from US
National Science Foundation grants number PHY-0070911 and PHY-95-28844.

%Tables:
\begin{table}[ht]
\caption{Branchings for proton captures on the most important waiting point
nuclei for different mass predictions from AW95 (AW) and this work SkX.
These branchings are the time
integrated averages obtained from our X-ray burst model.
\label{TabBranch}}
\begin{tabular}{ccccc} \hline
Waiting point & SkX & SkX-MIN--SkX-MAX & AW-MIN--AW-MAX & AW-MIN--AW-MAXEXP
\\ 
$^{60}$Zn & 95\% & 91\% - 97\% & 83\% - 98\% & 83\% - 99\% \\
$^{64}$Ge & 30\% & 0.5\% - 86\% & 0.0\% - 98\% & 0.0\% - 99\% \\
$^{68}$Se & 0.5\% & 0.0\% - 26\% & 0.0\% - 74\% & 0.0\% - 15\% \\
$^{72}$Kr & 0.0\% & 0.0\% - 8\%  & 0.0\% - 87\% & 0.0\% - 8\% \\ 
\end{tabular}
\end{table}

\begin{table}[ht]
\caption{Nuclei for which more a accurate mass would improve
the accuracy of rp process calculations in type I X-ray bursts. 
The upper part of the table
lists nuclei for which the current uncertainties lead to large
uncertainties in calculated burst time scales. The lower part of the
table lists nuclei, for which accurate masses are important, but
current estimates of the uncertainties do not lead to large
uncertainties in rp process calculations. Nevertheless, an
experimental confirmation for the masses being in the estimated
range would be important. 
Within each part, the nuclei are sorted by
uncertainty, so a measurement of the top ranked nuclei would be most important.
For each nucleus we list either the
experimental mass excess (Exp) (\protect\cite{b1} and 
\protect\cite{TBZ01} for $^{70}$Se)
or the
theoretical mass excess (SkX) calculated in this work in MeV.
\label{TabExp}}
\begin{tabular}{ccc} \hline
Nuclide & Exp  & SkX \\ 
$^{68}$Se & & -54.15 $\pm$ 0.30$^a$ \\
$^{64}$Ge & -54.43 $\pm$ 0.250 & \\
$^{70}$Kr & & -40.98 $\pm$ 0.16 \\
$^{70}$Se$^b$ & -61.60 $\pm$ 0.12 & \\
$^{65}$As & & -46.70 $\pm$ 0.14 \\
$^{69}$Br & & -46.13 $\pm$ 0.11 \\
$^{66}$Se & & -41.85 $\pm$ 0.10 \\
\hline
$^{72}$Kr & -54.11 $\pm$ 0.271 & \\
$^{73}$Rb & & -46.27 $\pm$ 0.17 \\
$^{73}$Kr$^b$ & -56.89 $\pm$ 0.14 & \\
$^{74}$Sr & & -40.67 $\pm$ 0.12 \\
$^{61}$Ga & & -47.14 $\pm$ 0.10 \\
$^{62}$Ge & & -42.38 $\pm$ 0.10 \\
\end{tabular}
$^a$ Theoretical estimate from AW95.\\
$^b$ Mirror to an rp process nucleus - a more accurate mass measurement could reduce
the error in the mass prediction for the proton rich mirror nucleus
by more than 30\%.
\end{table}

\bibliographystyle{prsty}

\newcommand{\noopsort}[1]{} \newcommand{\printfirst}[2]{#1}
  \newcommand{\singleletter}[1]{#1} \newcommand{\swithchargs}[2]{#2#1}

%\bibliography{sec_xrb}

\clearpage

\includegraphics[height=7in]{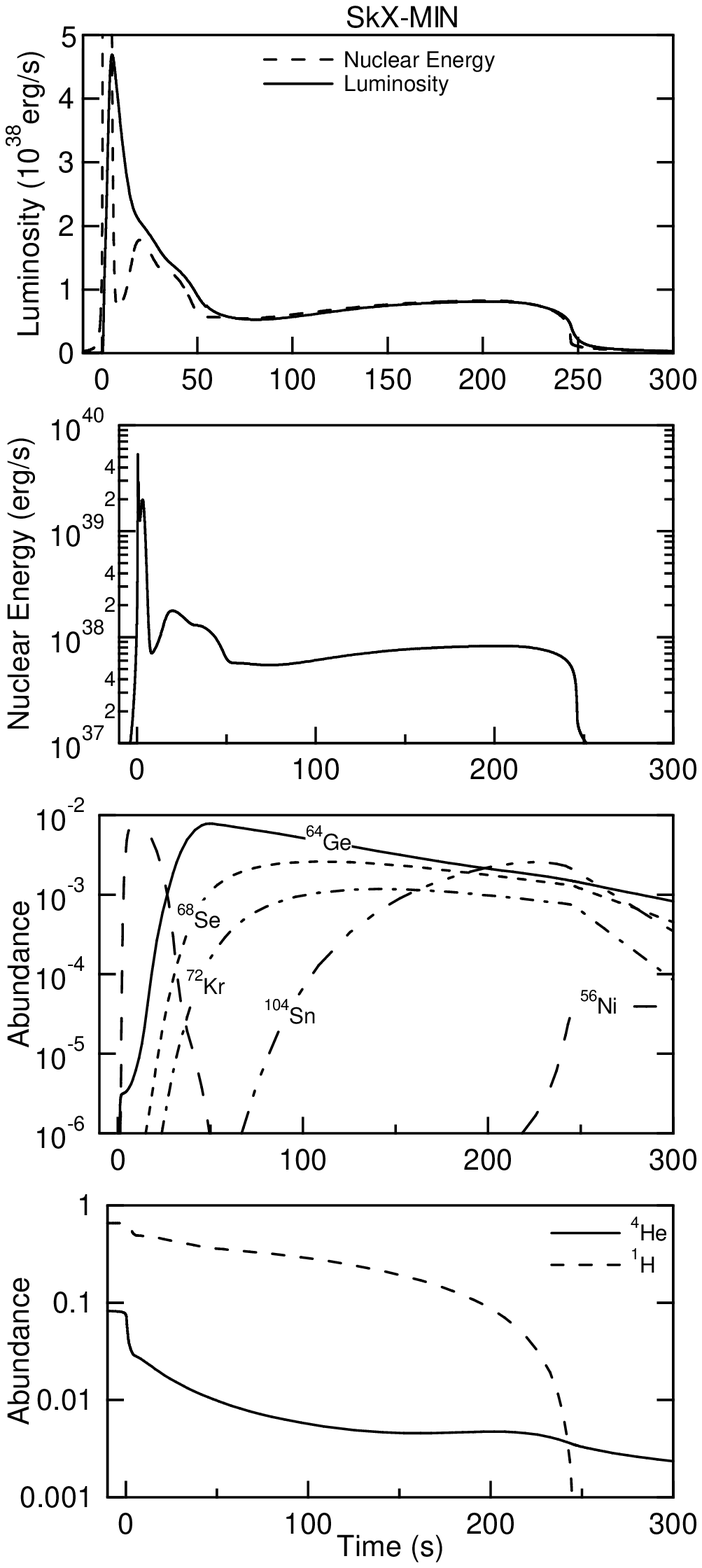}
\includegraphics[height=7in]{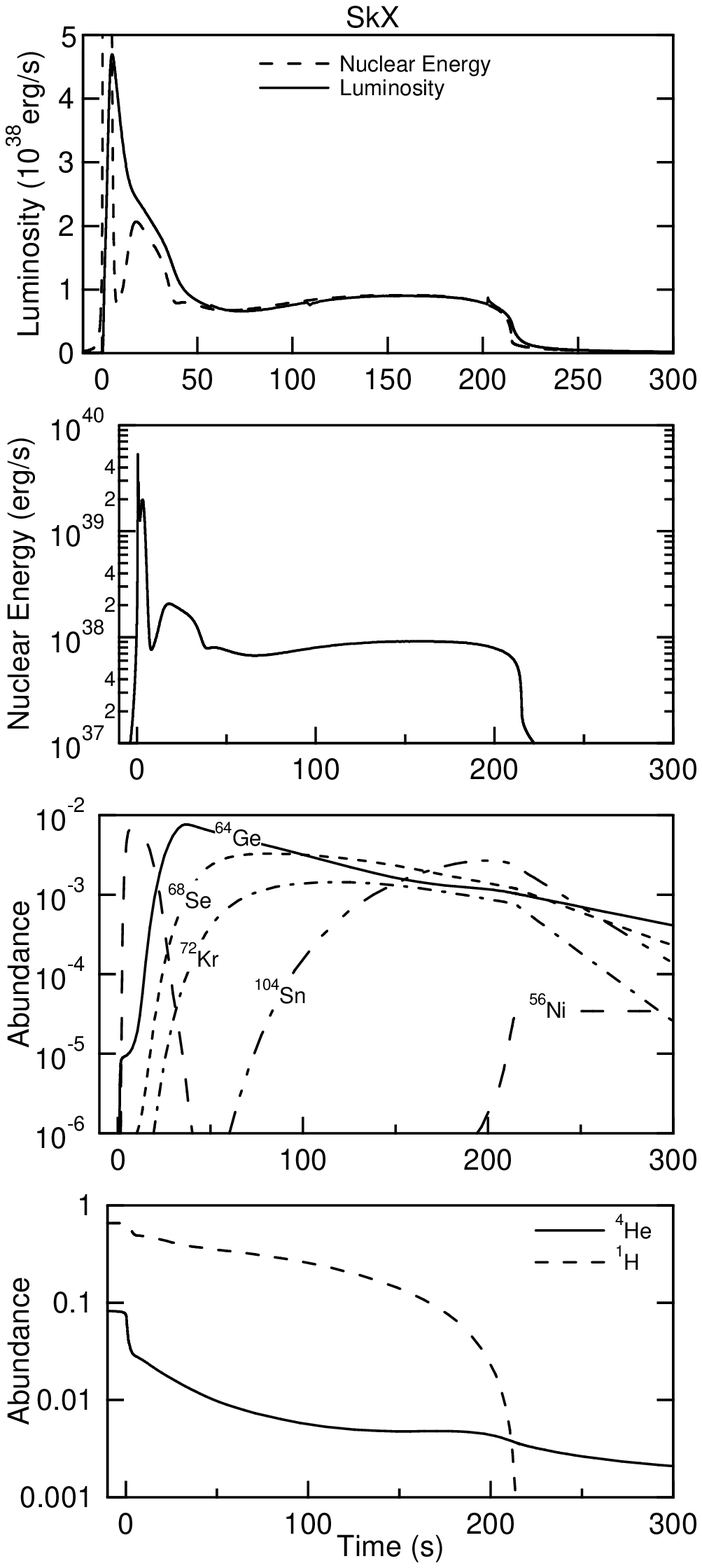}
\includegraphics[height=7in]{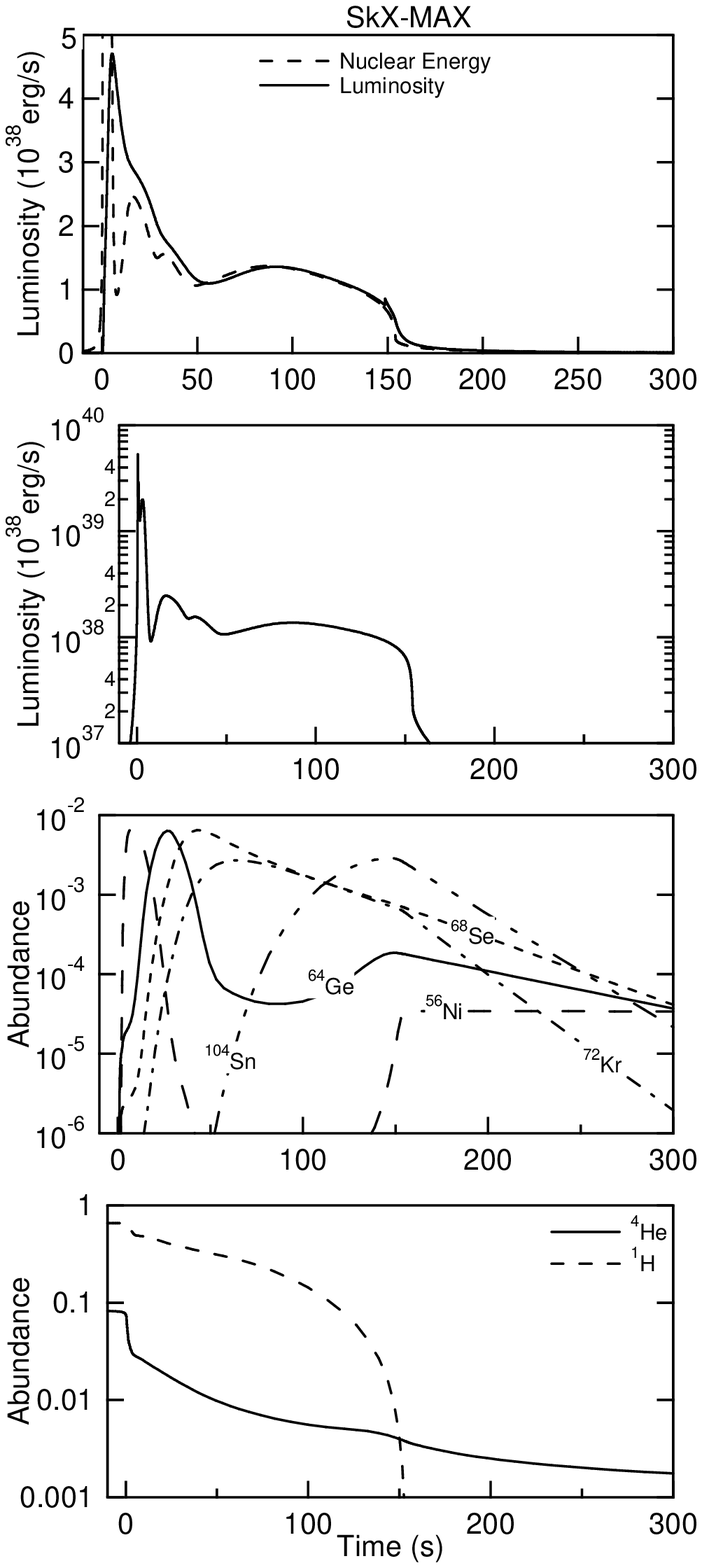}
\includegraphics[height=7in]{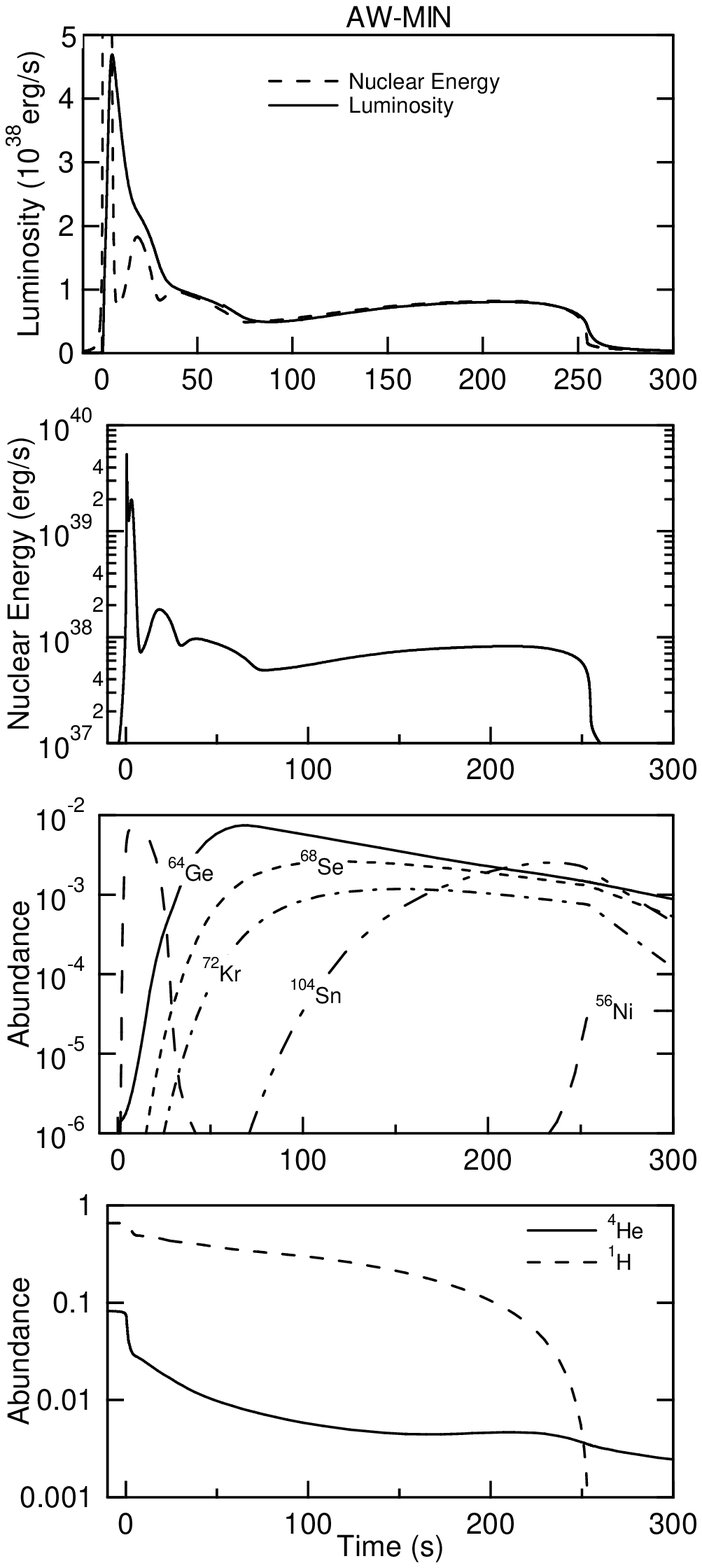}

\end{document}